\documentclass[aps,preprint]{revtex4}%
\usepackage{amsfonts}
\usepackage{amsmath}
\usepackage{amssymb}
\usepackage{graphicx}%
\begin{document}
\title{{\LARGE NATURAL ENTROPY PRODUCTION IN AN INFLATIONARY MODEL FOR A POLARIZED
VACUUM}}
\author{Marcelo Samuel Berman$^{1}$ and Murari M. Som$^{2}$ }
\affiliation{$^{1}$Instituto Albert Einstein \ - Av. Candido Hartmann, 575 - \ \# 17 \ -
80730-440 - Curitiba - PR - Brazil - E-mail: msberman@institutoalberteinstein.org}
\affiliation{$^{2}$Instituto de F\'{\i}sica da UFRJ - Ilha do Fundao, 21945-000, Rio de
Janeiro, RJ, Brazil}
\begin{abstract}
Though entropy production is forbidden in standard FRW Cosmology, Berman and
Som presented a simple inflationary model where entropy production by bulk
viscosity, during standard inflation without\ ad hoc pressure terms can be
accommodated with Robertson-Walker's metric, so the requirement that the early
Universe be anisotropic is not essential in order to have entropy growth
during inflationary phase, as we show. Entropy also grows due to shear
viscosity, for the anisotropic case. The intrinsically inflationary metric
that we propose can be thought of as defining a polarized vacuum, and leads
directly to the desired effects without the need of introducing extra pressure terms.

\end{abstract}
\maketitle

\begin{center}
{\LARGE NATURAL ENTROPY PRODUCTION IN AN INFLATIONARY MODEL FOR A POLARIZED
VACUUM}

\bigskip

Marcelo Samuel Berman and

Murari M. Som
\end{center}

\bigskip

\bigskip{\Large I. Introduction}

\bigskip\lbrack Note to the reader: this paper was originally prepared around
the year 1990. It was submitted in an earlier version, the referee asked for
clarifications, but for some reason, we made the clarifying in text, but did
not re-submit. As it seems to us that the contents is still adequate, we
publish it, at this time.]

Standard FRW Cosmology models do not provide a mechanism for entropy
production; unless one modifies the pressure by including an ad-hoc viscosity
term, like in Berman and Paim[6]. However, Berman and Som[1] presented a
special anisotropic model of the Bianchi type I that undergoes exponential
expansion and leads to entropy production directly, without further
assumptions. The geometry of the model during inflation is described by metric:

\bigskip

\bigskip$ds^{2}=-dt^{2}+A^{2}dx^{2}+B^{2}dy^{2}+C^{2}dz^{2}$ \ \ \ \ \ \ \ \ \ ,\ \ \ \ \ \ \ \ \ \ \ \ \ \ \ \ \ \ \ \ \ \ \ \ \ (1.1)

\bigskip

with

\bigskip

$A=A_{0}e^{D_{1}t}$ \ \ \ \ \ \ \ ,\ \ \ \ \ \ \ \ \ \ \ \ \ \ \ \ \ \ \ \ \ \ \ \ \ \ \ \ \ \ \ \ \ \ \ \ \ \ \ \ \ \ \ \ \ \ \ \ \ \ \ \ \ \ \ \ \ \ \ \ \ \ \ \ \ \ \ \ \ \ \ (1.2a)

$\bigskip$

$B=B_{0}e^{D_{2}t}$ \ \ \ \ \ \ ,\ \ \ \ \ \ \ \ \ \ \ \ \ \ \ \ \ \ \ \ \ \ \ \ \ \ \ \ \ \ \ \ \ \ \ \ \ \ \ \ \ \ \ \ \ \ \ \ \ \ \ \ \ \ \ \ \ \ \ \ \ \ \ \ \ \ \ \ \ \ \ \ (1.2b)

\bigskip

$C=C_{0}e^{D_{3}t}$ \ \ \ \ \ \ \ ,\ \ \ \ \ \ \ \ \ \ \ \ \ \ \ \ \ \ \ \ \ \ \ \ \ \ \ \ \ \ \ \ \ \ \ \ \ \ \ \ \ \ \ \ \ \ \ \ \ \ \ \ \ \ \ \ \ \ \ \ \ \ \ \ \ \ \ \ \ \ \ (1.2c)

\bigskip

where \ $A_{0},B_{0},C_{0}$\ \ and \ $D_{i}$\ 's are constants \ ($i=1,2,3$%
)\ \ \ . This metric can be thought of as representing a polarized vacuum.

\bigskip

The content of the model is the constant energy density \ $\rho$\ \ \ ,
isotropic constant pressure \ $p$\ \ \ and the trace-free constant anisotropic
pressure \ \ $\Pi_{i}^{k}$\ \ \ given by, in units \ \ \ $\frac{8\pi G}{c^{4}%
}=1$\ \ \ \ \ \ , as:

\bigskip

$\rho=D_{1}D_{2}+D_{2}D_{3}+D_{3}D_{1}$\ \ \ \ \ \ \ \ \ \ \ \ ,\ \ \ \ \ \ \ \ \ \ \ \ \ \ \ \ \ \ \ \ \ \ \ \ \ \ \ \ \ \ \ \ \ \ \ \ \ \ \ \ \ \ (1.3)

\bigskip

$p=-\frac{1}{3}\left[  \rho+2\left(  D_{1}^{2}+D_{2}^{2}+D_{3}^{3}\right)
\right]  $\ \ \ \ \ \ \ \ \ \ \ \ ,\ \ \ \ \ \ \ \ \ \ \ \ \ \ \ \ \ \ \ \ \ \ \ \ \ \ \ \ \ \ \ \ \ \ \ \ (1.4)

\bigskip

$\Pi_{i}^{i}=-3D\sigma_{i}^{i}$\ \ \ \ \ (no summation) \ \ \ \ \ \ \ \ \ \ \ \ \ \ \ \ \ \ \ \ \ \ \ \ \ \ \ \ \ \ \ \ \ \ \ \ \ \ \ \ \ \ \ \ \ \ \ \ (1.5)

\bigskip

where \ $\sigma_{i}^{i}$\ \ is the shearing tensor and \ $D=\frac{1}{3}\left(
D_{1}+D_{2}+D_{3}\right)  $\ . We see from (1.5) that the coefficient of
viscosity due to shear is a constant.

\bigskip

The negative isotropic pressure is the usual phenomenon associated with the
exponential expansion,

\bigskip

$H=$\ $\frac{1}{3}\left(  D_{1}+D_{2}+D_{3}\right)  =D$\ \ \ \ \ \ \ \ \ .

\bigskip

The conservation of thermal energy is then given by[2]:

\bigskip

$T\dot{S}=D\sigma_{k}^{i}\sigma_{i}^{k}$ \ \ \ \ \ \ \ \ \ ,

\bigskip

where \ $S$\ \ is the entropy density and \ \ $T$\ \ is the temperature. It
would be very attractive if we could explain the huge entropy per baryon,
denoted by the microwave background, by physical processes acting in the
early, or very early, Universe, as was remarked by Weinberg[3]. The present
work is a step towards that goal, and it will be shown that the requirement
that the Universe undergoes anisotropic inflation is not essential to entropy
production, because it does happen in the isotropic case, for isotropic inflation.

\bigskip

\bigskip{\Large II. The entropy production due to shear viscosity}

\bigskip

To obtain the entropy production, we define the entropy current according to
the second order dissipative relativistic fluid theories of Israel[4],
appropriate to this model,

\bigskip

$S^{\mu}=\frac{U^{\mu}}{T}\left[  TS-\frac{1}{2}\beta_{2}\Pi^{\alpha\beta}%
\Pi_{\alpha\beta}\right]  $ \ \ \ \ \ \ \ \ \ \ \ ,\ \ \ \ \ \ \ \ \ \ \ \ \ \ \ \ \ \ \ \ \ \ \ \ \ \ \ \ \ \ \ \ \ \ \ \ \ \ \ (2.1)

\bigskip

where \ $\beta_{2}$\ \ \ is the phenomenological coefficient, $S$\ \ \ is the
equilibrium entropy density and \ $U^{\mu}$\ \ is the 4-velocity,

\bigskip

$U^{\mu}U_{\mu}=1$ \ \ \ \ \ \ \ \ \ \ \ \ \ \ \ \ . \ \ \ \ \ \ \ \ \ \ \ \ \ \ \ \ \ \ \ \ \ \ \ \ \ \ \ \ \ \ \ \ \ \ \ \ \ \ \ \ \ \ \ \ \ \ \ \ \ \ \ \ \ \ \ \ \ \ \ \ \ \ (2.2)

\bigskip

Taking the divergence of (2.1), one shall obtain, supposing that the time
derivative of \ \ $\beta_{2}$\ \ is of first order,

\bigskip

\bigskip$S^{\mu};_{\mu}=-\frac{\Pi_{\alpha}^{\beta}}{T}\left(  U_{;\beta
}^{\alpha}+\beta_{2}\dot{\Pi}_{\beta}^{\alpha}\right)  $ \ \ \ \ \ \ , \ \ \ \ \ \ \ \ \ \ \ \ \ \ \ \ \ \ \ \ \ \ \ \ \ \ \ \ \ \ \ \ \ \ \ \ \ \ \ \ \ \ \ \ \ \ \ (2.3)

\bigskip

where \ $\dot{\Pi}_{\alpha\beta}=\Pi_{\alpha\beta_{;\lambda}}U^{\lambda}%
$\ \ \ , the semicolon denoting covariant derivative.

\bigskip

From (2.1), we find (2.3) by means of the following intermediate steps:

\bigskip

$S^{\mu}=\frac{U^{\mu}}{T}\left[  TS-\frac{1}{2}\beta_{2}\Pi^{\alpha\beta}%
\Pi_{\alpha\beta}\right]  =\frac{1}{T}\left[  pU^{\mu}-U_{\lambda}%
T^{\lambda\mu}\right]  -\frac{\beta_{2}}{2T}\Pi^{\alpha\beta}\Pi_{\alpha\beta
}U^{\mu}$\ \ \ \ \ .\ \ \ \ \ \ 

\bigskip

We now find:

\bigskip

$S_{;\mu}^{\mu}=-\frac{1}{T}U_{\lambda}T_{\text{ \ };\mu}^{\lambda\mu}+\left(
\frac{p}{T}U^{\mu}\right)  _{;\mu}-\left(  \frac{U_{\lambda}}{T}\right)
_{;\mu}T^{\lambda\mu}-\left[  \frac{1}{2T}\beta_{2}\Pi^{\alpha\beta}%
\Pi_{\alpha\beta}U^{\mu}\right]  _{;\mu}=$

$\bigskip$

$=-T^{-1}U_{\lambda}T_{\text{ \ };\mu}^{\lambda\mu}-T^{-1}\Pi^{\lambda\mu
}U_{\lambda;\mu}-\left[  \frac{1}{2T}\beta_{2}\Pi^{\alpha\beta}\Pi
_{\alpha\beta}U^{\mu}\right]  _{;\mu}$ \ \ \ \ \ \ \ \ \ \ . \ \ \ \ \ \ \ \ \ \ \ \ (2.3b)

\bigskip

In the case of interest to us, the energy momentum tensor is covariant divergence-less:

\bigskip

$T_{\text{ \ };\mu}^{\lambda\mu}=0$ \ \ \ \ \ \ \ \ \ \ .

\bigskip

In calculating the last term in (2.3b), we can treat \ $U_{\text{ };\mu}^{\mu
}$\ \ and the time derivatives of \ $\frac{\beta_{2}}{T}$\ , as the first
order quantities, since they vanish for equilibrium[7]. Hence, one obtains:

\bigskip

$TS_{;\mu}^{\mu}=-\Pi^{\alpha\beta}\left(  U_{\alpha;\beta}+\beta_{2}\dot{\Pi
}_{\alpha\beta}\right)  $\ \ \ \ \ \ .

\bigskip

Equation (2.3) gives the rate of entropy production per unity volume. The
second law of thermodynamics \ $\left(  S^{\mu};_{\mu}\geqslant0\right)
$\ \ will be satisfied identically if:

\bigskip

$\Pi_{\beta}^{\alpha}=-2n\left[  U^{\alpha};_{\beta}+\beta_{2}\dot{\Pi}%
_{\beta}^{\alpha}\right]  $ \ \ \ \ \ \ , \ \ \ \ \ \ \ \ \ \ \ \ \ \ \ \ \ \ \ \ \ \ \ \ \ \ \ \ \ \ \ \ \ \ \ \ \ \ \ \ \ \ \ \ \ \ \ \ \ \ (2.4)

\bigskip

where the positive coefficient of proportionality \ $n$ \ is the coefficient
of the shear viscosity.

\bigskip

From (1.5) we have \ $\dot{\Pi}_{\beta}^{\alpha}=0$\ \ \ , and $2n=3D=$%
\ \ constant. Thus in Bianchi I type model the entropy grows in a constant
rate. The coefficient of shear viscosity is \ $n=\frac{3}{2}H$\ .

\bigskip

During the period of inflation, the physical entropy is given by:

\bigskip

$S^{0}=S-\frac{\tau D^{2}}{T}\sigma_{\alpha\beta}$\ $\sigma^{\alpha\beta}%
$\ \ \ \ \ \ \ \ \ \ , \ \ \ \ \ \ \ \ \ \ \ \ \ \ \ \ \ \ \ \ \ \ \ \ \ \ \ \ \ \ \ \ \ \ \ \ \ \ \ \ \ \ \ \ \ \ \ \ \ \ \ \ \ \ (2.5)

\bigskip

where \ $S$\ \ is the equilibrium entropy density, and,

\bigskip

$\tau=\frac{9}{2}\beta_{2}$\ \ \ \ \ \ \ \ . \ \ \ \ \ \ \ \ \ \ \ \ \ \ \ \ \ \ \ \ \ \ \ \ \ \ \ \ \ 

\bigskip

{\Large III. Entropy production in the isotropic case}

\bigskip

Let us consider the case of a small anisotropy such as:

\bigskip

$D_{1}=D_{0}+\alpha$ \ \ \ \ \ \ ,

\bigskip

$D_{2}=D_{0}+\beta$ \ \ \ \ \ \ ,

\bigskip

$D_{3}=D_{0}+\gamma$ \ \ \ \ \ \ ,

\bigskip

where \ $\alpha,\beta$\ \ and \ \ $\gamma$\ \ \ are small quantities.

\bigskip

From (1.3) and (1.4) one obtains, neglecting the second order terms,

\bigskip

$\rho\cong3D_{0}^{2}+6D_{0}(\alpha+\beta+\gamma)$ \ \ \ \ \ , \ \ \ \ \ \ \ \ \ \ \ \ \ \ \ \ \ \ \ \ \ \ \ \ \ \ \ \ \ \ \ \ \ \ \ \ \ \ \ \ \ \ \ \ \ \ \ (3.1)

\bigskip

$p=-\frac{1}{3}\left[  9D_{0}^{2}+10D_{0}(\alpha+\beta+\gamma)\right]  $
\ \ \ \ . \ \ \ \ \ \ \ \ \ \ \ \ \ \ \ \ \ \ \ \ \ \ \ \ \ \ \ \ \ \ \ \ \ \ \ \ \ (3.2)

\bigskip

From (3.1) and (3.2) one finds that:

\bigskip

\bigskip$p=-3D_{0}^{2}-\frac{10}{3}D_{0}(\alpha+\beta+\gamma)$ \ \ \ \ \ \ \ , \ \ \ \ \ \ \ \ \ \ \ \ \ \ \ \ \ \ \ \ \ \ \ \ \ \ \ \ \ \ \ \ \ \ \ \ \ \ \ \ \ (3.3)

\bigskip

up to the first order.

\bigskip

In the isotropic exponential expansion one has:

\bigskip

\bigskip$p=-3D_{0}^{2}=-\rho$ \ \ \ \ \ \ \ \ \ , \ \ \ \ \ \ \ \ \ \ \ \ \ \ \ \ \ \ \ \ \ \ \ \ \ \ \ \ \ \ \ \ \ \ \ \ \ \ \ \ \ \ \ \ \ \ \ \ \ \ \ \ \ \ \ \ \ (3.4)

\bigskip

which is the usual vacuum state condition for the inflationary model.

\bigskip

If we impose the condition \ $p=-\rho$\ \ , then relation\ (1.4) yields:

\bigskip

$D_{1}^{2}+D_{2}^{2}+D_{3}^{2}=D_{1}D_{2}+D_{2}D_{3}+D_{3}D_{1}$\ \ \ \ \ \ \ \ \ \ \ \ ,\ \ \ \ \ \ \ \ \ \ \ \ \ \ \ \ \ \ \ \ \ \ \ \ (3.5)

\bigskip

leading to $D_{1}=D_{2}=D_{3}$\ \ \ . However, for a vacuum state cosmology in
a Bianchi-I space-time, one needs, apart \ from a shear viscosity, a bulk
viscosity given by:

\bigskip

$\Pi=-\frac{2}{3}\left[  D_{1}D_{2}+D_{2}D_{3}+D_{3}D_{1}+D_{1}+D_{2}%
+D_{3}\right]  $ \ \ \ \ \ . \ \ \ \ \ \ \ \ \ \ \ (3.6)

\bigskip

This leads to the modification of (2.1). The entropy current takes the form:

\bigskip

$S^{\mu}\approx SU^{\mu}-\frac{1}{2}\left[  \beta_{0}\Pi^{2}+\beta_{2}%
\Pi^{\alpha\beta}\Pi_{\alpha\beta}\right]  \frac{U^{\mu}}{T}$
\ \ \ \ \ \ \ \ \ , \ \ \ \ \ \ \ \ \ \ \ \ \ \ \ \ \ \ \ \ \ \ \ \ (3.7)

\bigskip

where $\beta_{0}$\ \ is a phenomenological coefficient. The inclusion of the
bulk viscosity further augments the rate of production of entropy in the
general case.

\bigskip

{\Large IV. Conclusions}

\bigskip

The requirement that the early Universe be anisotropic is not essential in
order to have entropy growth during inflation. We showed that entropy grows,
due to bulk and shear viscosity, in the anisotropic case. An estimate of the
maximum rate at which the entropy of the primordial plasma increases in the
presence of a homogeneous shear, and the minimum entropy increase, were
calculated by Vischniac[5]; who also discussed the physical mechanism
underlying our proposed shear viscosity, which can be achieved through a
massive scalar field interacting with a massive gauge field. The interest in
our model resides in the natural way of entropy production, without adding
ad-hoc viscosity terms to the pressure; all one has to do is define the vacuum
polarization metric (1.1) and apply Einstein's equations.

\bigskip\ \ \ 

{\Large Acknowledgements}

\bigskip

One of the authors (MSB) gratefully thanks his intellectual mentors, Fernando
de Mello Gomide and M. M. Som (which is the second author), and is also
grateful for the encouragement by Geni, Albert, and Paula. An anonymous
referee saved us from a typing mistake in one of the formulae, and we thank
him vehemently.

\bigskip

{\Large References}

\bigskip

1. Berman,M.S.; Som, M.M. (1989) - GRG \textbf{21}, 967.

\bigskip

2. Ellis, GRE (1971) - Proc. S.I.F., Course XLVII, edited by R. Sachs,
Academic Press, New York.

3. Weinberg, S. (1972) - \textit{Gravitation and Cosmology}, Wiley, N.Y.

\bigskip

4. Israel, W. (1976) - Ann. Phys.(N.Y.) \textbf{100}, 259.

\bigskip

5. Vishniac, E.T. (1983) - MNRAS \textbf{205}, 675-681.

\bigskip

6. Berman,M.S.; Paim, T. (1990) - Nuovo Cimento \textbf{105B}, 1377.

\bigskip

7. Israel, W. (1976) - Ann. Phys.(N.Y.) \textbf{100}, 310-331.

\end{document}